\documentclass[11pt]{article}
\pdfoutput=1
\usepackage{jheppub}
\usepackage{amsthm}

\usepackage{hyperref}
\usepackage{xcolor}
\usepackage{mathtools}
\usepackage{tensor}
\usepackage{empheq}

\usepackage{tikz}
\usepackage{graphicx}
\usepackage{subfigure}
\usetikzlibrary{calc,fadings,decorations.pathreplacing}
\usepackage{verbatim}
\usepackage[normalem]{ulem}
\usepackage{multirow}
\usepackage{makecell}

\DeclareMathOperator{\Tr}{Tr}

\preprint{LCTP-25-03}

\title{The Superconformal Index and Black Hole Instabilities}

\author[a]{Evan Deddo}

\author[a,b]{Leopoldo A. Pando Zayas}

\author[a]{and Wenjie Zhou}

\emailAdd{evdedd@umich.edu, lpandoz@umich.edu, vincezh@umich.edu}

\affiliation[a]{Leinweber Center for Theoretical Physics, 
University of Michigan, Ann Arbor, MI 48109, USA}

\affiliation[b]{The Abdus Salam International Centre for Theoretical Physics, 34014 Trieste, Italy}

\abstract{The  superconformal index  of ${\cal N}=4$ supersymmetric Yang-Mills theory with gauge group $\mathrm{U}(N)$ has provided powerful insights into the entropy of supersymmetric black holes in AdS$_5\times S^5$, including some sub-leading logarithmic and non-perturbative corrections. Recently, the phase space of supersymmetric solutions has been argued to contain configurations other than the asymptotically AdS$_5$ black hole. Such configurations include the so-called grey galaxies where the black hole at the center is surrounded by a gas of gravitons. By numerically evaluating the superconformal index  of ${\cal N}=4$ supersymmetric Yang-Mills at small values of $N$, we detect systematic deviations from the entropy of black holes with two distinct angular momenta. We find that the giant graviton expansion of the index is a numerically efficient way of evaluating the index that complements the direct character evaluation and allows for explicit access to $N\le 15$ with up to two giant gravitons in the expansion. We find it remarkable that a supersymmetric quantity in field theory, usually thought of as a rigid counting observable, indeed contains information about different phases in the space of supersymmetric solutions on the gravity side.}

\begin{document}

\maketitle

\section{Introduction}

Providing a statistical foundation for black hole entropy is one of the central tests for any candidate theory of quantum gravity. In the context of the AdS/CFT correspondence, where the counting of degrees of freedom can be outsourced to the field theory side, this herculean labor has been completed with the help of a certain supersymmetric partition function.  The superconformal index (SCI) of four-dimensional ${\cal N}=4$ supersymmetric Yang-Mills has provided a microscopic foundation for the entropy of the dual black holes in AdS$_5\times S^5$ \cite{Cabo-Bizet:2018ehj,Choi:2018hmj,Benini:2018ywd}. 

Beyond a microscopic foundation for the Bekenstein-Hawking entropy, that is, the quarter of the area of the black hole, the SCI has provided significant insight into quantum corrections to the entropy. The first study of sub-leading corrections, such as terms logarithmic in area, was presented in \cite{GonzalezLezcano:2020yeb}. These logarithmic corrections to the entropy were later directly corroborated in the field theory side in  \cite{Amariti:2020jyx, Amariti:2021ubd} and given an effective field theory interpretation in \cite{Cassani:2021fyv,ArabiArdehali:2021nsx}. On the gravitational side, these logarithmic corrections arise as the result of massless particles running in quantum loops. Agreement between these field-theoretic logarithmic corrections and near-horizon based approaches was presented in  \cite{David:2021qaa}. Higher-curvature corrections, which arise when Einstein gravity is viewed as an effective field theory,  were successfully addressed in \cite{Melo:2020amq,Bobev:2022bjm,Cassani:2022lrk} for example. Beyond logarithmic and higher-curvature corrections to the entropy of the dual black holes, the superconformal index has even allowed for the interpretation of certain non-perturbative contributions \cite{Aharony:2021zkr,Chen:2023lzq,Cabo-Bizet:2023ejm}.
    
The SCI has also been able to shed light on some of the various gravitational phases that are expected from supersymmetric solutions to the equations of motion. For example, certain properties of the confinement/deconfinement transition in field theory were discussed in  \cite{Choi:2018vbz,Cabo-Bizet:2019eaf,ArabiArdehali:2019orz,Copetti:2020dil}.

In this manuscript we explore the ability of the index to indicate the existence of phases other than the black hole. We are largely motivated by recent works arguing for specific instabilities of rotating AdS black holes \cite{Kim:2023sig,Choi:2024xnv,Bajaj:2024utv}. In particular, we are interested in testing the hypothesis of contributions from grey galaxies—  configurations in which the central black hole is surrounded by a graviton gas in a manner compatible with the conserved quantities.

To explore this question, we numerically evaluate the SCI for finite values of the gauge group rank $N$ and read off its degeneracies. Our main goal is to compare the index degeneracies with the entropy of the corresponding black hole to show that there is a systematic discrepancy.  Technically, we utilized previous precedents of numerically evaluating the index  \cite{Agarwal:2020zwm, Murthy:2020scj} following its definition. Other discussions utilizing direct numerical evaluation at small rank include \cite{Lezcano:2021qbj,Benini:2021ano}, a more recent revision of the Bethe-Ansatz approach \cite{Cabo-Bizet:2024kfe}. We also pursue a numerical evaluation of the index by means of the giant graviton expansion.

The SCI has proven to be an extremely versatile tool for probing quantum aspects of black holes. In this manuscript, the SCI is used to explore the configuration space of supersymmetric solutions. We hope that this lends support to the idea that it can be explored to elucidate more intricate quantum aspects of gravity.

The rest of the note is organized as follows. In section \ref{Sec:SCI} we review the SCI and introduce the precise degeneracies we will compute. Section \ref{Sec:GGE} describes the giant graviton expansion of the index, adapted to the needs of this paper. Section \ref{Sec:SCI-v-BH} compares the degeneracy of states extracted from the SCI with the entropy of the semiclassical black holes. We conclude and provide additional outlook in Section \ref{Sec:Conclusions}.

{\bf Note added:} While we were readying our results for posting, we received \cite{Choi:2025lck} with a comprehensive discussion of the index vis-\`a-vis the space of supersymmetric solutions.


\section{The superconformal index}\label{Sec:SCI}
In this section, we define a partially refined version of the superconformal index that will be studied numerically.
Recall that states in $\mathcal{N}=4$ SYM of $S^1 \times S^3$ may be labeled by their energy $E$, angular momenta $J_1$, $J_2$ on $S^3$, and $\mathrm{SO}(6)$ R-charges $Q_1$, $Q_2$, $Q_3$. The full superconformal index is defined as a Witten index \cite{Kinney:2005ej}
\begin{equation}\label{eq:full_index_trace}
    \mathcal{I}(p,q,y_a)=\Tr(-1)^F e^{-\beta\{\mathcal{Q},\mathcal{Q}^\dagger\}} p^{J_1}q^{J_2}y_1^{Q_1}y_2^{Q_2}y_3^{Q_3},
\end{equation}
where $\mathcal{Q}$ is one of the 16 supercharges, and the fugacities satisfy $pq=y_1 y_2 y_3$. Due to cancellations from the $(-1)^F$ factor, only BPS states with $E=J_1+J_2+Q_1+Q_2+Q_3$ contribute to the index. (Different choices for the supercharge $\mathcal{Q}$ lead to different relative signs in this BPS constraint, but as a convention we choose the unique supercharge that yields all plus signs.) The index may be computed by plethystically exponentiating the index for single trace operators ($\mathcal{I}_\text{single}$) and projecting to gauge singlets. Explicitly \cite{Agarwal:2020zwm},
\begin{equation}\label{eq:index_char}
    \mathcal{I}(p,q,y_a) =\oint d \mu_{\mathfrak{g}}(\mathbf{z}) \operatorname{PE}\left[I_{\text {single }}\left(p,q,y_a\right) \chi_{\mathfrak{g}}(\mathbf{z})\right],
\end{equation}
where $d \mu_{\mathfrak{g}}(\mathbf{z})$ and $\chi_{\mathfrak{g}}(\mathbf{z})$ are the Harr measure and adjoint character for the gauge algebra $\mathfrak{g}$. For gauge group $\mathrm{U}(N)$ we have

\begin{equation*}
    \oint d \mu_{\mathfrak{g}}(\mathbf{z}) = \oint_{|z_1|=1}\!\!\!\!\!\cdots\oint_{|z_N|=1}\frac{dz_1}{2\pi i z_1}\ldots \frac{dz_N}{2\pi i z_N}\prod_{\underset{j>k}{j,k=1}}^N\left(1-\frac{z_j}{z_k}\right),
\end{equation*}
\begin{equation}\label{eq:U-N-char}
    \chi_{\mathfrak{g}}(\mathbf{z}) = \sum_{j,k=1}^N\frac{z_j}{z_k},
\end{equation}
and the full single letter index for $\mathcal{N}=4$ SYM is
\begin{equation}
    I_{\text {single }}\left(p,q,y_a)\right) = 1-\frac{\left(1-y_1\right)\left(1-y_2\right)\left(1-y_3\right)}{(1-p)(1-q)}.
\end{equation}
However, in this work, we will restrict our attention to unrefined versions of $I_{\text {single }}$ that will be described later.

Various limits of the index may be taken by imposing constraints on the fugacities. The unrefined index with only a single independent fugacity $x$ was studied numerically in \cite{Agarwal:2020zwm,Murthy:2020scj}. There, the fugacities were set to $p=q=x^3$ and $y_1=y_2=y_3=x^2$. In terms of grand canonical ensemble averages, this limit restricts to equal angular momentum and equal charge. Microcanonically, however, states of all angular momenta and charges are still included in the counting. The index reduces to
\begin{equation}\label{eq:single_fug}
    \mathcal{I}(x)=\Tr(-1)^F e^{-\beta\{\mathcal{Q},\mathcal{Q}^\dagger\}} x^{6(J_L+Q)},
\end{equation}
where $J_{L,R} = (J_1\pm J_2)/2$ and $Q=(Q_1+Q_2+Q_3)/3$. The degeneracy at each level $j\equiv6(J_L+Q)$ was found to correlate closely with the entropy of supersymmetric black holes in $AdS_5\times S^5$.

\subsection{Unequal angular momenta}
In this work we consider the more general case of unequal angular momentum $J_1\neq J_2$. We construct a more refined version of the above index by first performing a fugacity scaling
\begin{equation}\label{eq:fug scaling}
    p\to px^3, \quad q\to qx^3, \quad y_a\to y_a x^2, 
\end{equation}
which preserves the product relation $pq=y_1 y_2 y_3$. We then have
\begin{equation}\label{eq:x index}
    \mathcal{I}(x,p,q,y_a)=\Tr(-1)^F x^{6(J_L+Q)}p^{J_1}q^{J_2}y_1^{Q_1}y_2^{Q_2}y_3^{Q_3}.
\end{equation}
(The $e^{-\beta\{\mathcal{Q},\mathcal{Q}^\dagger\}} $ has been omitted with the understanding that the trace is only over BPS states.) Since there is now extra redundancy among the fugacities, we may freely choose the constraint $pq=y_1 y_2 y_3=1$. Note that, in this context, the previously defined single-fugacity index corresponds to $p=q=y_1=y_2=y_3=1$. More generally, we now consider $p$ unconstrained. The relation $q=1/p$ replaces $p^{J_1}q^{J_2}\to p^{2J_R}$ so that the index tracks the difference between angular momentum $J_1$ and $J_2$. One may restrict to equal charges in two ways: (i) by setting $y_1=y_2=y_3=1$ just as in \cite{Agarwal:2020zwm,Murthy:2020scj}, and (ii) microcanonically, by discarding individual states with unequal charges. In both cases, the resulting index will label states by only $j\equiv 6(J_L+Q)$ and $J_R$.

\subsection{The equal-fugacity constraint}\label{sec:=fugacity}
Applying the equal fugacity constraint $y_1=y_2=y_3=1$ to \eqref{eq:x index} yields
\begin{equation}\label{eq:px index}
    \mathcal{I}(x,p)=\Tr(-1)^F x^j p^{2J_R},
\end{equation}
and the corresponding single letter index reduces to
\begin{equation}\label{eq:px single}
    1-\frac{\left(1-x^2\right)^3}{(1-px^3)(1-x^3/p)}.
\end{equation}

In this limit, the index still counts all states, including those with unequal charges. However, since the chemical potentials for the charges have been tuned equal, the index is now fully symmetric in $Q_i$. Therefore, ensemble averages for the $Q_i$ are equal when viewed in the context of the canonical ensemble. This two-fugacity limit of the index will be our main object of study.

\subsection{The equal-charge constraint}
Alternatively, one may consider a different charge restriction in which states of unequal charge are explicitly discarded. If we at first impose no constraints on the $y_a$'s other than $y_1 y_2 y_3=1$, we have
\begin{equation}
    \mathcal{I}(x,p,y_1,y_2)=\Tr(-1)^F x^j p^{2J_R}y_1^{Q_1-Q_3}y_2^{Q_2-Q_3},
\end{equation}
and the single letter index is
\begin{equation}
    1-\frac{\left(1-y_1 x^2\right)\left(1-y_2 x^2\right)\left(1-\frac{x^2}{y_1 y_2}\right)}{(1-px^3)(1-x^3/p)}.
\end{equation}
The restriction to equal charges cannot be done on the single-letter index, since non-equal charge letters may combine to form equal-change states. (Based on the trace formula, a naive attempt to restrict the index to equal charges would be $\lim y_1\to 0$, $\lim y_2\to 0$. However, this method cannot work because $Q_1-Q_3$ and $Q_2-Q_3$ may take either sign.)

Therefore, one must proceed by computing the index via the plethystic exponential, expanding in terms of $x,p,y_1$ and $y_2$, and only afterwards discard any terms containing $y_1$ and $y_2$. However, expanding the index in all four fugacities is extremely computationally intensive. In this current work, we will not consider the equal-charge constraint further.

\section{The giant graviton expansion of the index}\label{Sec:GGE}

The giant graviton expansion is a particular expansion of the index. It has been recently discussed for the superconformal index in \cite{Gaiotto:2021xce,Imamura:2021ytr}, inspired by previous developments in the context of the Schur index \cite{Bourdier:2015wda,Bourdier:2015sga}. The generic form of giant graviton expansions is 
\begin{equation}\label{eq:GGE}
    \frac{{\cal I}_{N}(p,q,y_a)}{{\cal I}_{\infty}(p,q,y_a)}=1+\sum\limits_{m=1}^\infty G_N^{(m)}(p,q,y_a),
\end{equation}
where $G_N^{(m)}(p,q,y_a)$ is the contribution from $m$ giant gravitons.

A matrix model explanation of the giant graviton expansion for the index was recently provided in \cite{Murthy:2022ien}. The authors of \cite{Liu:2022olj} presented a study comparing the matrix model approach with Immamura's expansion in \cite{Imamura:2021ytr} which is based on the structure of the intersecting branes index. Further clarifications on numerical evaluation and the nature of analytic continuations were presented in \cite{Ezroura:2024wmp}.

Giant graviton expansions have powerful implications in the context of the AdS/CFT correspondence, as they can also be interpreted geometrically on the gravity side. Indeed, a holographic reproduction of the giant graviton expansion of the $\frac{1}{2}$-BPS index using probe D3 branes was reported in \cite{Lee:2023iil,Eleftheriou:2023jxr}. Subsequent insights from the fully back-reacted geometry were discussed in \cite{Chang:2024zqi} and \cite{Deddo:2024liu}.
    
In this manuscript we pursue the giant graviton expansion as a practical way to evaluate the SCI. In particular, we will determine the degeneracies from the giant graviton expansion for the index with two charges $(j, J_R)$.

In \eqref{eq:GGE} we make use of the multigraviton index

\begin{eqnarray} \label{eq:multigraviton}
    {\cal I}_\infty (p,q,y_a)&=& \prod\limits_{k=1}^\infty \frac{1}{1-I_{\rm single}(p,q,y_a)}\nonumber \\
    &=& \prod\limits_{k=1}^\infty \frac{(1-p^k)(1-q^k)}{(1-y_1^k)(1-y_2^k)(1-y_3^k)},
\end{eqnarray}

and concentrate on the giant graviton expansion as described in \cite{Liu:2022olj} following \cite{Murthy:2022ien}:

\begin{equation}\label{eq:giant_term}
    G_N^{(m)}(\mathbf{g})=\left[\frac{(-1)^m}{m!}\left(\prod_{i=1}^m \frac{w_i / z_i}{1-w_i / z_i}\right) \operatorname{det}\left(\frac{1}{1-w_j / z_i}\right)_{i, j=1}^m \exp \left(-\sum_{k=1}^{\infty} \frac{\gamma_k}{k} \alpha_k \beta_k\right)\right]_{w_i^{-N} z_i^N},
\end{equation}



\begin{equation}\label{eq:alphabeta}
    \alpha_k=\sum_{i=1}^m z_i^k-w_i^k, \quad \quad \beta_k=\sum_{i=1}^m z_i^{-k}-w_i^{-k}, \quad \quad
    \gamma_k=\frac{(1-p^k)(1-q^k)}{(1-y_1^k)(1-y_2^k)(1-y_3^k)}-1.
\end{equation}
The subscript at the end of \eqref{eq:giant_term} instructs us to single out one particular term in the $w_i$, $z_i$ expansion.
As in Section \ref{sec:=fugacity}, we consider the limit
\begin{equation}
    p \to px^3, \quad q\to x^3/p, \quad y_a=x^2.
\end{equation}

Using \eqref{eq:GGE}-\eqref{eq:alphabeta}
we may compute the index and expand the result as a polynomial in $x$ and $p$:
\begin{equation}
    \mathcal{I}_N(x,p)=\sum_{j,J_R} d(j,J_R)x^jp^{2J_R}.
\end{equation}
As we compute terms at larger $j$, we must include an increasing number of giant graviton terms. The contribution from the $m$-th giant enters at \cite{Liu:2022olj}
\begin{equation}
    j=2m(N+m).
\end{equation}
Since the number of auxiliary parameters $w_i,z_i$ grows as $2m$, the computational resources required to expand and evaluate $G_N^{(m)}$ increase dramatically. However, for large enough $N$ the giant graviton expansion can be significantly more efficient than direct evaluation of \eqref{eq:index_char}, as long as one stays within the appropriate range of $j$. The direct expression involving characters requires an expansion in $N$ fugacities, whereas the number of $w_i$ and $z_i$ at fixed giant number is independent of $N$.


\section{Index versus black hole entropy and the $J_R$ instability}\label{Sec:SCI-v-BH}

\subsection{Black hole entropy}\label{sec:BH_entropy}

Let us briefly review the entropy of the supergravity solutions presented in \cite{Gutowski:2004ez, Gutowski2004, Kunduri2006} whose full non-supersymmetric version is \cite{Chong:2005hr}. Further clarifications on taking the supersymmetric limit of the black holes and the precise relation to the dual field theory were presented in  \cite{Cabo-Bizet:2018ehj}. 

For supersymmetric AdS$_5$ black holes with equal electric charges $Q=Q_1=Q_2=Q_3$, the entropy and nonlinear charge constraint are
\begin{equation}\label{eq:S}
S=2 \pi \sqrt{3Q^2-N^2 J_L},
\end{equation}
and
\begin{equation}\label{Eq:Constraint}
    Q^3+\frac{N^2}{2}(J_L^2-J_R^2) = \left(\frac{N^2}{2}+3Q\right)\left(3Q^2-N^2J_L\right).
\end{equation}
On the supergravity side, this constraint is associated with the condition of avoiding closed timelike curves in the background  \cite{Chong:2005hr}. In field theory, the charge constraint arises by imposing that the entropy, as computed from the SCI defined with complex chemical potentials,  has no imaginary part \cite{Choi:2018hmj}. Some recent discussion on the nature of the constraint can be found in \cite{Larsen:2021wnu,Larsen:2024fmp}. In the presence of higher curvature terms, there are also subleading corrections to the constraint \cite{Cassani:2022lrk}. As we will briefly review, the constraint plays an important role in determining the phase space of supergravity solutions.

In section \ref{sec:index_v_BH} we will compare the black hole entropy $S(j,J_R)$ at equal $R$-charge with the degeneracy of the index \eqref{eq:px index}. To express \eqref{eq:S} in terms of $j$ and $J_R$, we must first solve the nonlinear charge constraint to find $J_L$ in terms of $j$ and $J_R$. The result is

\begin{equation}
    J_L(j,J_R)=-\frac{\left(\mathcal{F}(j,J_R)^{1/3}-2 N^2\right) \left(\mathcal{F}(j,J_R)^{1/3}-2 \left(2 j+N^2\right)\right)}{24 \mathcal{F}(j,J_R)^{1/3}},
\end{equation}
where
\begin{align}
    \mathcal{F}(j,J_R)=
    4 N^2 \bigg(3 j^2&+6 j N^2-108 J_R^2+2 N^4\nonumber\\
    &-\sqrt{4 j N^2 \left(j^2-324 J_R^2\right)+9 \left(j^2-36 J_R^2\right)^2-432 J_R^2 N^4}\; \bigg).
\end{align}
From \eqref{eq:S} the entropy is then
\begin{equation}\label{eq:S(j,J_R)}
    S(j,J_R)=2\pi\sqrt{3\left(\frac{j}{6}-J_L(j,J_R)\right)^2-N^2J_L(j,J_R)}.
\end{equation}

Let us now consider the allowed range of $J_R$ under the condition $S>0$. From \eqref{eq:S(j,J_R)}, the entropy is positive when
\begin{equation}
    J_L\leq\frac{1}{6} \left(-\sqrt{2 j N^2+N^4}+j+N^2\right).
\end{equation}
Substituting this result into \eqref{Eq:Constraint} and solving for $J_R$ yields
\begin{equation}
    |J_R|\leq\frac{1}{6} \sqrt{j^2+\frac{2}{3} j N \left(3 N-2 \sqrt{2 j+N^2}\right)+\frac{2}{3} N^3 \left(N-\sqrt{2 j+N^2}\right)}.
\end{equation}

At fixed $j$, the entropy is maximum at $J_R=0$ and decreases to zero on either side at the maximal value of $|J_R|$. The function $S(j,J_R)$ is plotted in Fig \ref{fig:S(j,J_R}. 

\begin{figure}[h!]
    \centering
    \includegraphics[width=0.6\linewidth]{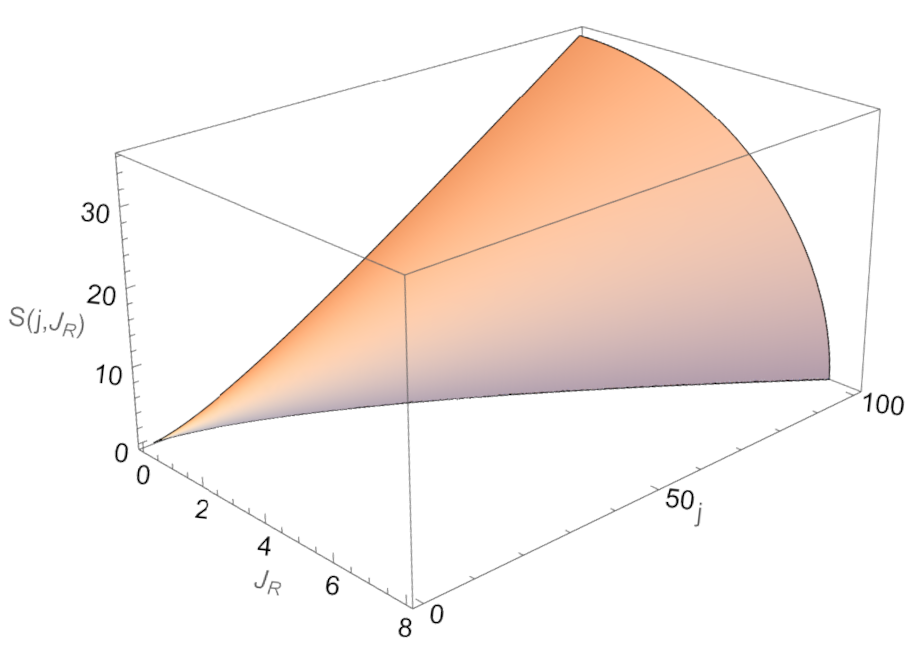}
    \caption{$S(j,J_R)$ for $N=10$. To compare this entropy to the single-charge index, we restrict to the $J_R$=0 plane. For the two-charge index studied in Section \ref{sec:index_v_BH} we plot cross sections at fixed $j$.}
    \label{fig:S(j,J_R}
\end{figure}

\subsection{Index versus black hole entropy for one charge $j$}\label{sec:single charge}

In this Subsection we reproduce the results of \cite{Agarwal:2020zwm,Murthy:2020scj}. This is more of a calibration of our numerical methods. We also point out some details that will be relevant for our main computation, which considers two charges. 

The expression \eqref{eq:index_char} can be used to quickly evaluate \eqref{eq:single_fug} for small $N$. In Mathematica, we designed an efficient algorithm tailored to $N\leq 4$ that utilizes the symmetry between the $z_i$'s in \eqref{eq:index_char} and pre-computes subcomponents of the index before finally assembling them (see Appendix \ref{app:numerical}). Expanding the index to $j=100$ requires about 0.3 seconds for $N=2$ and less than a minute for $N=4$. However, for $N\geq 5$ this algorithm becomes intractable, and one must move to a more brute force expansion approach. Due to the excessive computation time required, we in fact avoid using the character method for the $N=10,15$ results later in the paper and instead utilize the giant graviton expansion.

\begin{figure*}[h]
    \centering
    \includegraphics[width = 0.8\textwidth]{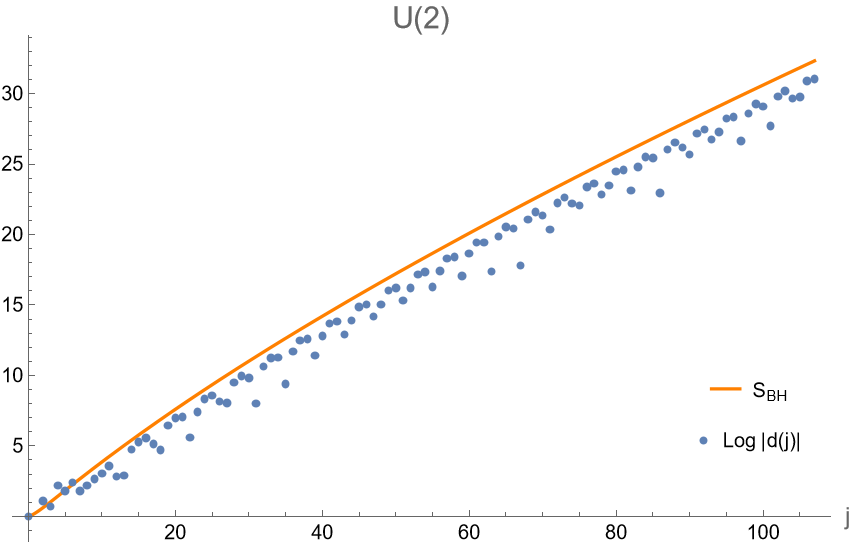}
    \caption{Degeneracy of the single-fugacity index at N=2 using the character method.}
    \label{fig:N=2_single}
\end{figure*}

Fig. \ref{fig:N=2_single} shows the degeneracy of the $\mathrm{U}(2)$ index, where $d(j)$ refers to the coefficient of $x^j$ in the expansion of the index:
\begin{equation*}
    \mathcal{I}_2(x)=\sum_{j=0}^\infty d(j)x^j= 1+3x^2-2x^3+9x^4-6x^5+11x^6-6x^7+9x^8+14x^9-21x^{10}+\cdots
\end{equation*}

The orange line represents the black hole entropy \eqref{eq:S(j,J_R)} at $J_R=0$. After an initial excess around $j\lesssim5$, the index follows the entropy curve closely, although from somewhat below. Due to the sign-alternating nature of the index, the plot contains roughly periodic dips where there are large cancellations between states.

In our study of the two-fugacity index in Section \ref{sec:index_v_BH}, we will plot the degeneracy as a function of $J_R$ with $j$ fixed. To make the most meaningful comparison to the black hole entropy, we specifically choose values of $j$ that avoid the aforementioned cancellations (i.e. values of $j$ where the plotted degeneracy is closest to the orange entropy curve in Fig. \ref{fig:N=2_single}).

\subsubsection{Giant graviton evaluation}

Estimates of the black hole entropy using the giant graviton expansion have been presented in \cite{Choi:2022ovw,Beccaria:2023hip} using the large $N$ limit via certain saddle-point approximations. In this note we will follow a direct numerical evaluation for $N\geq10$.

Let us first make some comments that follow from \cite{Murthy:2022ien} and the subsequent explicit numerical implementation in \cite{Liu:2022olj}. The contribution for the $m$-th giant graviton, $G_N^{(m)}$, appears at the minimal charge $j_{\rm min}=2m(N+m)$. This expression is numerically verified for $N=2$ in Table 2 of \cite{Liu:2022olj}. In particular, the finite-$N$ index exactly matches the multi-graviton index $\mathcal{I}_\infty$ up until the $j$ at which the first giant starts contributing, $j_{\rm min} (m=1)=2(N+1)$. (See also Table 1 of \cite{Murthy:2020scj}.) Similarly, the single giant term alone is sufficient up until $j_\mathrm{min}(m=2)=4(N+2)$, and two giants are sufficient up until $j_\mathrm{min}(m=3)=6(N+3)$, etc.

Although the one-giant term signals the first clear point of departure from the multi-graviton index, it does not necessarily describe black hole contributions reliably. To enter the black hole regime, we would like to have $j\sim N^2$. This requires giant graviton contributions of the order of $m_*: N^2\sim 2m_*(N+m_*)$ with $m_*\sim \frac{\sqrt{3}-1}{2}\, N$. For $N=15$, this result suggests that $m_*\sim 6$ would be required to put us in the regime where black hole states contribute. However, our numerical results below will demonstrate that $m=2$ already begins to show high agreement between the index and black hole entropy.

\begin{figure}[htb]
    \centering
    \includegraphics[width=0.8\linewidth]{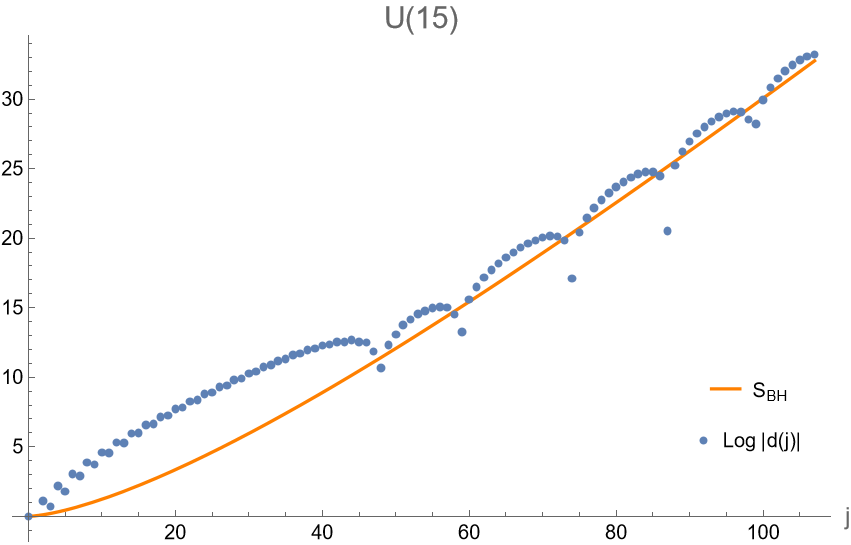}
    \caption{Degeneracy of the single-fugacity index at N=15 using the two-giant-graviton approximation.}
    \label{fig:N=15_single}
\end{figure}

As a sanity check, and to develop a precise sense of the resources required, we first reproduce the one-charge results presented in  \cite{Agarwal:2020zwm, Murthy:2020scj}. In Figure \ref{fig:N=15_single} we plot the black hole entropy (solid line) versus the index degeneracy for $N=15$. We use the two-giant-graviton expansion and therefore continue up to the maximum reliable value of $j_\mathrm{min}(m=3)-1=107$. This computation requires far fewer resources than those needed to reach this value of $N$ in the direct character evaluation. Using \eqref{eq:index_char} would require expanding the index in 15 variables, whereas the $m=2$ giant graviton expansion requires only 4. The latter method is therefore exponentially faster, and requires roughly no more resources than the character method required at $\mathrm{U}(4)$. Accordingly, Figure \ref{fig:N=15_single} was produced in less than one minute.

Despite the previous estimation that 6 giants would be needed to reach the black hole regime, Figure \ref{fig:N=15_single} already begins to show high agreement with the entropy curve. Although increasing to $j=151$ by including three giants would be desirable, this would require performing an expansion in 6 variables that precludes use of our more efficient algorithm. This computation would in fact require similar resources as $\mathrm{U}(6)$ at $j=151$, which becomes intractable with a brute force expansion.

\subsection{Index versus black hole entropy for $(j,J_R)$}\label{sec:index_v_BH}

We now turn to our main object of study, the two-fugacity index \eqref{eq:px index}. As before, index may be computed via the character method in \eqref{eq:index_char}, but we now expand in both $x$ and $p$ to compute the degeneracy $d_N(j,J_R)$. For $N=2$ we find

\begin{align}\label{eq:I_2(x,p)}
    \mathcal{I}_2(x,p)&=\sum_{j=0}^\infty d(j,J_R)x^j p^{2J_R}\nonumber\\
    &=1+3x^2-(p^{-1}+p^1)x^3+9x^4-(3p^{-1}+3p)x^5+(-p^{-2}+13-p^2)x^6\nonumber\\
    &\hspace{0.7cm}-(3p^{-1}+3p)x^7+(-3p^{-2}+15-3p^2)x^8+(7p^{-1}+7p)x^9\nonumber\\
    &\hspace{0.7cm}+\cdots\nonumber\\
    &\hspace{0.7cm}+(8p^{-6}+309p^{-4}+985p^{-2}+1500+985p^2+309p^4+8p^6)x^{24}\nonumber\\
    &\hspace{0.7cm}+\cdots
\end{align}
From this expansion, one may observe the qualitative similarity between the degeneracy of the index and the entropy $S(j,J_R)$ in \eqref{eq:S(j,J_R)}. At each particular $j$, the $p$-series multiplying $x^j$ contains a finite number of terms and is symmetric in the power of $p$. This behavior matches the discussion in section \ref{sec:BH_entropy}, since the range of allowed $J_R$ is finite and $S(j,J_R)$ is symmetric with respect to the sign of $J_R$. As $j$ increases, the allowed range of $J_R$ increases and the length of the corresponding $p$-series grows.

Expanding the index directly in both $x$ and $p$ is slower than the computation for the single-fugacity index. For large $j$, it actually becomes more efficient to use a modified approach in Mathematica. Rather than expand in $p$, we first choose a fixed numerical value for $p$ in the single-letter index \eqref{eq:px single}. This reduces the problem back to a much faster single-fugacity computation. We now view \eqref{eq:I_2(x,p)} as a series in $x$ with variable coefficients that depend on the choice of $p$. After calculating the coefficients for sufficiently many values of $p$, the exact series can be determined using a polynomial fit.

We may leverage certain structural constraints of the $p$-series to significantly simplify the fitting procedure. Along with symmetry under \( p^j \to 1/p^j \), the parity of the maximum power of $p$ aligns with the parity of \( j \); that is, for even (odd) \( j \), the polynomial also possesses an even (odd) degree. 
Despite requiring multiple evaluations of \( p \), the fitting method remains computationally efficient because single-fugacity evaluations are considerably faster than direct expansion of the full index. This observation is consistent with the fact that performing the plethystic exponentiation for each term is already computationally intensive, and the introduction of an additional variable \( p \) further exacerbates the complexity.

\begin{figure}[htb]
\subfigure[$\mathrm{U}(2)$, $j=99$]
{
    \begin{minipage}{7.5cm}
    \centering
    \includegraphics[width=0.8\linewidth, keepaspectratio=true]{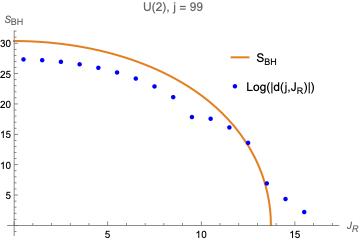}
    \end{minipage}
}
\subfigure[$\mathrm{U}(2)$, $j=200$]
{
    \begin{minipage}{7.5cm}
    \centering
    \includegraphics[width=0.8\linewidth, keepaspectratio=true]{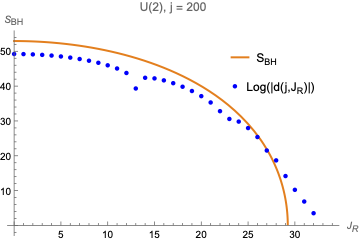}
    \end{minipage}
}
\subfigure[$\mathrm{U}(3)$, $j=99$]
{
    \begin{minipage}{7.5cm}
    \centering
    \includegraphics[width=0.8\linewidth, keepaspectratio=true]{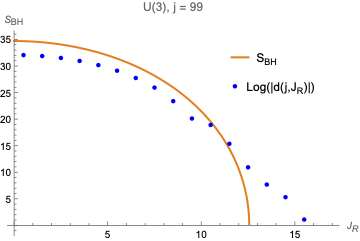}
    \end{minipage}
}
\subfigure[$\mathrm{U}(3)$, $j=200$]
{
    \begin{minipage}{7.5cm}
    \centering
    \includegraphics[width=0.8\linewidth, keepaspectratio=true]{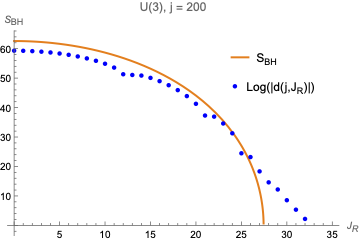}
    \end{minipage}
}
\subfigure[$\mathrm{U}(4)$, $j=98$]
{
    \begin{minipage}{7.5cm}
    \centering
    \includegraphics[width=0.8\linewidth, keepaspectratio=true]{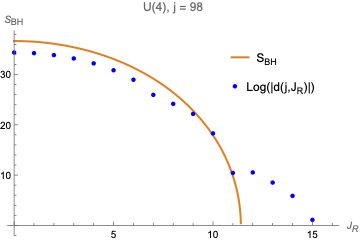}
    \end{minipage}
}
\subfigure[$\mathrm{U}(4)$, $j=199$]
{
    \begin{minipage}{7.5cm}
    \centering
    \includegraphics[width=0.8\linewidth, keepaspectratio=true]{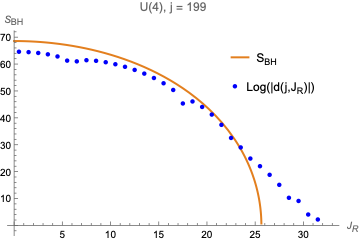}
    \end{minipage}
}
\caption{The black hole entropy and the numerically-computed index entropy for $\mathrm{U}(2),\mathrm{U}(3), $ and $\mathrm{U}(4)$.}\label{fig:U(2)-U(4)}
\end{figure}

Some results of these computations are shown in Figure \ref{fig:U(2)-U(4)}. For $\mathrm{U}(2)$, $\mathrm{U}(3)$, and $\mathrm{U}(4)$ we selected fixed values of $j$ and computed the degeneracy $d_N(j,J_R)$ as a function of $J_R$. As before, the log of the degeneracy (blue points) is plotted along with the entropy $S(j,J_R)$ of the corresponding black hole solution (orange curve.)

Unlike the case for the single-charge index in \ref{sec:single charge}, the two-charge index shows a clear deviation from the black hole entropy in the region of large $J_R$. For $\mathrm{U}(4)$ with $j=98$, for example, the degeneracy is still non-zero up to $J_R=15$ even though $S_{BH}$ reaches zero at $J_R\sim11.5$. The point of divergence occurs at $J_R=11$. All of the other graphs show qualitatively similar behavior. Moreover, by comparing the results at $j\sim100$ and $j\sim200$, we demonstrate that the discrepancy is somewhat persistent with variable $j$. As $j$ increases, the blue "tails" incorporate an increasing number of points but only experience minor variations in shape.

\subsubsection{Giant graviton evaluation for $(j, J_R)$}

\begin{figure}[htb]
\centering 
\subfigure[$\mathrm{U}(10)$, $j=77$]
{
    \begin{minipage}{7.4cm}
    \centering
    \includegraphics[width=0.9\linewidth, keepaspectratio=true]{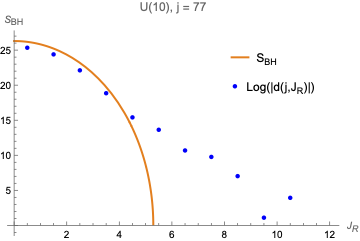}
    \end{minipage}
}
\subfigure[$\mathrm{U}(15)$, $j=107$]
{
    \begin{minipage}{7.4cm}
    \centering
    \includegraphics[width=0.9\linewidth, keepaspectratio=true]{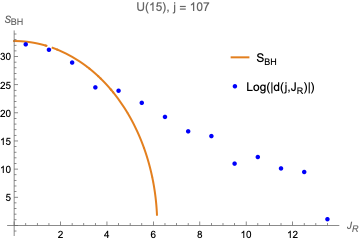}
    \end{minipage}
}
\caption{Numerically computed indicial entropy using the giant graviton approach, evaluated at $\mathrm{U}(10)$, $j=77$ and $\mathrm{U}(15)$, $j=107$.}\label{fig:U(10)U(15)}
\end{figure}

We now utilize the giant graviton expansion to explore a regime of larger $N$. We carry out the expansion to $m=2$, as discussed previously. Although there is no firm barrier limiting the value of $N$ in this approach, there is a fixed ceiling on $j$ corresponding to the point where the third giant contribution becomes necessary. To access the black hole regime, we therefore restrict to $N=10$ and $N=15$, since we previously verified a consistency between the single-charge $\mathrm{U}(15)$ index and the black hole entropy. Figure \ref{fig:U(10)U(15)} compares the index degeneracy with $S_{BH}$, just as in Figure \ref{fig:U(2)-U(4)}.

The plots show a clear divergence beyond a critical value of $J_R$, supporting the existence of a phase transition between the black hole and a grey galaxy phase. The tails have also become more pronounced. This behavior suggests that the length of the tail may increase even further at higher $N$, although we leave the investigation of such questions to future work.

\section{Conclusions}\label{Sec:Conclusions}

In this manuscript we have evaluated the SCI numerically following the standard definition based on the character expansion. We have also complemented this approach, in a certain regime,  with an evaluation based on the giant graviton expansion. The giant graviton expansion provides access to a complementary region and supports the robustness of the systematic deviation between the microcanonical index and the black hole entropy as we increase the rank of the gauge group $N$. In particular, within the range where the value of the index is given by the two-giant-graviton contribution we are able to explore $N=15$ up to $j_{\rm max}= 107$. In the case of the index for two distinct angular momenta, we found a systematic deviation from the entropy of a single-center AdS black hole. We established that this deviation persists as we increase the value of $j$  as well as $N$. Therefore, we confirmed that the superconformal index detects contributions to the entropy that are beyond those provided by the supersymmetric black hole asymptoting to AdS$_5\times S^5$. Some of the recently proposed configurations that could contribute to the index are the grey galaxies wherein the black hole at the center is surruounding by a supersymmetric gas of gravitons. The ability of the superconformal index to detect different phases in the space of supersymmetric gravity solutions is quite impressive and opens the door to the possibility of exploring aspects of gravity in the very quantum regime of small values of $N$.

The phase space of supersymmetric solutions in AdS$_4$ has recently been discussed in a series of papers \cite{Kim:2023sig,Choi:2024xnv}. It would be interesting to find support for the structure of the phase space of solutions by analyzing the superconformal index of the dual field theory on $S^1\times S^2$. Evaluation of the SCI in the large$-N$ limit leads to results that agree with the entropy of the dual black hole \cite{Choi:2019zpz,Nian:2019pxj}.  Unfortunately, direct numerical evaluation of the SCI is technically more challenging in this case due to the sum over non-perturbative contributions in the form of monopoles. It would be interesting to explore alternative techniques, such as those employed in \cite{Amariti:2023ygn} which rely on certain universal properties of the coefficients of the R-symmetry or flavor symmetry 2-point current correlation functions. 
        
More broadly, it would be interesting to explore to which extent the potential instabilities discussed in this manuscript are universal properties of quantum gravity in asymptotically AdS spacetimes. Given that the SCI in various dimensions provides a microscopic explanation for the entropy of large asymptotically AdS black holes, it could serve as a guide for exploring different phases. 
        
Small black holes in AdS resemble those in asymptotically flat spacetimes, since they both possess negative specific heat. It would be interesting to extract information relevant for the thermodynamics of small black holes using the superconformal index by revisiting early attempts \cite{Choi:2021lbk}. Finally, if we take the superconformal index as truly the Phythia of quantum gravity, that is, if the index contains information about all phases, we should use it as a tool to learn about quantum gravity in the regime where there might not be a semi-classical description in terms of supergravity configurations.

\section*{Acknowledgments}
We are thankful to Jim Liu for suggesting the use of giant graviton expansions to numerically evaluate the SCI and other valuable comments. We thank Vineeth Krishna for discussions in the very early stages of this project. The authors thank Elena Rasia for a discussion that led to a numerical idea that reduced the two-charge index to a one-charge computation. This work is partially supported by the U.S. Department of Energy under grant DE-SC0007859. ED is supported in part by a Leinweber Graduate Fellowship.

\appendix
\section{Discussion of numerical methods}\label{app:numerical}

In this appendix, we summarize the method used in Sections \ref{sec:single charge} and \ref{sec:index_v_BH} to compute the SCI at $N\leq 4$. We also compare the relative efficiency of various numerical approaches for the computation of the SCI depending on the rank $N$ and the charge $j$.

To implement equations \eqref{eq:index_char}-\eqref{eq:U-N-char} efficiently, we change to an alternative (but equivalent \cite{Agarwal:2020zwm}) form of the measure,

\begin{equation}\label{eq:new_measure}
    \oint d \mu_{\mathfrak{g}}(\mathbf{z}) = \frac{1}{N!}\oint\prod_{\ell=1}^N\frac{dz_\ell}{2\pi i z_\ell}\prod_{\underset{j\neq k}{j,k=1}}^N\left(1-\frac{z_j}{z_k}\right),
\end{equation}
and rewrite \eqref{eq:index_char} as
\begin{equation}\label{eq:trick1}
    \mathcal{I}=\frac{1}{N!}\text{PE}[NI_\text{single}]\oint\prod_{\ell=1}^N\frac{dz_\ell}{2\pi i z_\ell}\prod_{\underset{i>k}{i,k=1}}^N \left\{\left(1-\frac{z_i}{z_k}\right)\left(1-\frac{z_k}{z_i}\right)\text{PE}\left[I_\text{single}\left(\frac{z_i}{z_k}+\frac{z_k}{z_i}\right)\right]\right\}.
\end{equation}
After expanding $I_\text{single}$ as a series in $x$ up to $\mathcal{O}(x^{j_\text{max}})$, the term in braces may be reorganized as
\begin{equation}\label{eq:trick2}
    \left(1-\frac{z_i}{z_k}\right)\left(1-\frac{z_k}{z_i}\right)\text{PE}\left[I_\text{single}\left(\frac{z_i}{z_k}+\frac{z_k}{z_i}\right)\right]\quad\to\quad\sum_{j=-\bar\jmath}^{\bar\jmath}c_j(x)\left(\frac{z_i}{z_k}\right)^j,
\end{equation}
where $\bar\jmath=\operatorname{Floor}(j_\text{max}/2)+1$. (Note that the expression cannot be expanded in $z_i/z_k$ first due the appearance of both positive and negative powers in the exponential.) The $c_j(x)$ are independent of $i,k$ and each only needs to be computed once. Also, by using \eqref{eq:new_measure} we have gained the useful symmetry property $c_j(x)=c_{-j}(x)$. After taking the product in \eqref{eq:trick1}, only terms where all the $z_i$ cancel contribute to the integral. This reduces the index to a sum of combinations of $c_j(x)$ satisfying certain algebraic relations. For example, the $N=4$ index is a sum of terms of the form
\begin{equation}\label{eq:trick3}
    c_{j_1}(x)c_{j_2}(x)\cdots c_{j_6}(x),
\end{equation}
where $j_1+j_2+j_3=0$, $j_1-j_4-j_5=0$ and $j_2+j_4-j_6=0$. This procedure is very efficient for $N\leq 4$, although it quickly becomes intractable for larger $N$.

Table \ref{tab:comparison} compares reasonable values of $j_\text{max}$ that one may achieve using this approach (column 1), the methods of \cite{Agarwal:2020zwm, Murthy:2020scj,Choi:2025lck}, as well as the giant graviton expansion. Note that carrying out \eqref{eq:trick1}-\eqref{eq:trick3} required only $\sim5.5$ minutes for $N=4$, $j=200$ so it could presumably be used to explore higher in $j$ as well. The methods of \cite{Agarwal:2020zwm, Murthy:2020scj,Choi:2025lck} were not reproduced for this study and we simply quote their results. The method relying on characters of the symmetric group was carried out explicitly for $N=10$, $j=70$ in \cite{Murthy:2020scj} and up to $N=10$, $j=90$ in \cite{Choi:2025lck}. However, once the necessary characters are calculated the index may be computed at various $N$ with a comparatively small amount of additional computation time \cite{Murthy:2020scj}.
\begin{table}[h]
\centering
\begin{tabular}{c||c|c|c|c}
$N$ & \eqref{eq:trick1}-\eqref{eq:trick2} & \makecell{Direct expansion\\ \cite{Agarwal:2020zwm}} & \makecell{$S_d$ characters \\ \cite{Murthy:2020scj,Choi:2025lck}} & GGE ($m=2$) \\ \hline\hline
$2$ & \multirow{3}{*}{200+} & \multirow{4}{*}{100} & \multirow{11}{*}{90} & 29 \\
$3$ & & & & 35 \\
$4$ & & & & 41 \\ \cline{2-2}
$5$ & \multirow{8}{*}{-} & & & 47 \\ \cline{3-3}
$6$ & & 70 & & 53 \\ \cline{3-3}
$7$ & & \multirow{6}{*}{-} & & 59 \\
$\cdots$ & & & & $\cdots$ \\
$12$ & & & & 89 \\
$13$ & & & & 95 \\
$14$ & & & & 101 \\
$15$ & & & & 107 \\
\end{tabular}
\caption{Achievable values of $j_\text{max}$ at a range of $N$ using various methods.}
\label{tab:comparison}
\end{table}

\bibliographystyle{JHEP}
\bibliography{step1ref}

\providecommand{\href}[2]{#2}\begingroup\raggedright\begin{thebibliography}{10}

\bibitem{Cabo-Bizet:2018ehj}
A.~Cabo-Bizet, D.~Cassani, D.~Martelli and S.~Murthy, \emph{{Microscopic origin of the Bekenstein-Hawking entropy of supersymmetric AdS$_{5}$ black holes}}, \href{https://doi.org/10.1007/JHEP10(2019)062}{\emph{JHEP} {\bfseries 10} (2019) 062}, [\href{https://arxiv.org/abs/1810.11442}{{\ttfamily 1810.11442}}].

\bibitem{Choi:2018hmj}
S.~Choi, J.~Kim, S.~Kim and J.~Nahmgoong, \emph{{Large AdS black holes from QFT}},  \href{https://arxiv.org/abs/1810.12067}{{\ttfamily 1810.12067}}.

\bibitem{Benini:2018ywd}
F.~Benini and E.~Milan, \emph{{Black Holes in 4D $\mathcal{N}$=4 Super-Yang-Mills Field Theory}}, \href{https://doi.org/10.1103/PhysRevX.10.021037}{\emph{Phys. Rev. X} {\bfseries 10} (2020) 021037}, [\href{https://arxiv.org/abs/1812.09613}{{\ttfamily 1812.09613}}].

\bibitem{GonzalezLezcano:2020yeb}
A.~Gonz\'alez~Lezcano, J.~Hong, J.~T. Liu and L.~A. Pando~Zayas, \emph{{Sub-leading Structures in Superconformal Indices: Subdominant Saddles and Logarithmic Contributions}}, \href{https://doi.org/10.1007/JHEP01(2021)001}{\emph{JHEP} {\bfseries 01} (2021) 001}, [\href{https://arxiv.org/abs/2007.12604}{{\ttfamily 2007.12604}}].

\bibitem{Amariti:2020jyx}
A.~Amariti, M.~Fazzi and A.~Segati, \emph{{The SCI of $ \mathcal{N} $ = 4 USp(2N$_{c}$) and SO(N$_{c}$) SYM as a matrix integral}}, \href{https://doi.org/10.1007/JHEP06(2021)132}{\emph{JHEP} {\bfseries 06} (2021) 132}, [\href{https://arxiv.org/abs/2012.15208}{{\ttfamily 2012.15208}}].

\bibitem{Amariti:2021ubd}
A.~Amariti, M.~Fazzi and A.~Segati, \emph{{Expanding on the Cardy-like limit of the SCI of 4d $ \mathcal{N} $ = 1 ABCD SCFTs}}, \href{https://doi.org/10.1007/JHEP07(2021)141}{\emph{JHEP} {\bfseries 07} (2021) 141}, [\href{https://arxiv.org/abs/2103.15853}{{\ttfamily 2103.15853}}].

\bibitem{Cassani:2021fyv}
D.~Cassani and Z.~Komargodski, \emph{{EFT and the SUSY Index on the 2nd Sheet}}, \href{https://doi.org/10.21468/SciPostPhys.11.1.004}{\emph{SciPost Phys.} {\bfseries 11} (2021) 004}, [\href{https://arxiv.org/abs/2104.01464}{{\ttfamily 2104.01464}}].

\bibitem{ArabiArdehali:2021nsx}
A.~Arabi~Ardehali and S.~Murthy, \emph{{The 4d superconformal index near roots of unity and 3d Chern-Simons theory}}, \href{https://doi.org/10.1007/JHEP10(2021)207}{\emph{JHEP} {\bfseries 10} (2021) 207}, [\href{https://arxiv.org/abs/2104.02051}{{\ttfamily 2104.02051}}].

\bibitem{David:2021qaa}
M.~David, A.~Lezcano~Gonz\'alez, J.~Nian and L.~A. Pando~Zayas, \emph{{Logarithmic corrections to the entropy of rotating black holes and black strings in AdS$_{5}$}}, \href{https://doi.org/10.1007/JHEP04(2022)160}{\emph{JHEP} {\bfseries 04} (2022) 160}, [\href{https://arxiv.org/abs/2106.09730}{{\ttfamily 2106.09730}}].

\bibitem{Melo:2020amq}
J.~a.~F. Melo and J.~E. Santos, \emph{{Stringy corrections to the entropy of electrically charged supersymmetric black holes with $\mathrm{AdS}_5\times S^5$ asymptotics}}, \href{https://doi.org/10.1103/PhysRevD.103.066008}{\emph{Phys. Rev. D} {\bfseries 103} (2021) 066008}, [\href{https://arxiv.org/abs/2007.06582}{{\ttfamily 2007.06582}}].

\bibitem{Bobev:2022bjm}
N.~Bobev, V.~Dimitrov, V.~Reys and A.~Vekemans, \emph{{Higher derivative corrections and AdS5 black holes}}, \href{https://doi.org/10.1103/PhysRevD.106.L121903}{\emph{Phys. Rev. D} {\bfseries 106} (2022) L121903}, [\href{https://arxiv.org/abs/2207.10671}{{\ttfamily 2207.10671}}].

\bibitem{Cassani:2022lrk}
D.~Cassani, A.~Ruip\'erez and E.~Turetta, \emph{{Corrections to AdS$_{5}$ black hole thermodynamics from higher-derivative supergravity}}, \href{https://doi.org/10.1007/JHEP11(2022)059}{\emph{JHEP} {\bfseries 11} (2022) 059}, [\href{https://arxiv.org/abs/2208.01007}{{\ttfamily 2208.01007}}].

\bibitem{Aharony:2021zkr}
O.~Aharony, F.~Benini, O.~Mamroud and E.~Milan, \emph{{A gravity interpretation for the Bethe Ansatz expansion of the $\mathcal{N}=4$ SYM index}}, \href{https://doi.org/10.1103/PhysRevD.104.086026}{\emph{Phys. Rev. D} {\bfseries 104} (2021) 086026}, [\href{https://arxiv.org/abs/2104.13932}{{\ttfamily 2104.13932}}].

\bibitem{Chen:2023lzq}
Y.~Chen, M.~Heydeman, Y.~Wang and M.~Zhang, \emph{{Probing supersymmetric black holes with surface defects}}, \href{https://doi.org/10.1007/JHEP10(2023)136}{\emph{JHEP} {\bfseries 10} (2023) 136}, [\href{https://arxiv.org/abs/2306.05463}{{\ttfamily 2306.05463}}].

\bibitem{Cabo-Bizet:2023ejm}
A.~Cabo-Bizet, M.~David and A.~Gonz\'alez~Lezcano, \emph{{Thermodynamics of black holes with probe D-branes}}, \href{https://doi.org/10.1007/JHEP06(2024)193}{\emph{JHEP} {\bfseries 06} (2024) 193}, [\href{https://arxiv.org/abs/2312.12533}{{\ttfamily 2312.12533}}].

\bibitem{Choi:2018vbz}
S.~Choi, J.~Kim, S.~Kim and J.~Nahmgoong, \emph{{Comments on deconfinement in AdS/CFT}},  \href{https://arxiv.org/abs/1811.08646}{{\ttfamily 1811.08646}}.

\bibitem{Cabo-Bizet:2019eaf}
A.~Cabo-Bizet and S.~Murthy, \emph{{Supersymmetric phases of 4d $ \mathcal{N} $ = 4 SYM at large $N$}}, \href{https://doi.org/10.1007/JHEP09(2020)184}{\emph{JHEP} {\bfseries 09} (2020) 184}, [\href{https://arxiv.org/abs/1909.09597}{{\ttfamily 1909.09597}}].

\bibitem{ArabiArdehali:2019orz}
A.~Arabi~Ardehali, J.~Hong and J.~T. Liu, \emph{{Asymptotic growth of the 4d $ \mathcal{N} $ = 4 index and partially deconfined phases}}, \href{https://doi.org/10.1007/JHEP07(2020)073}{\emph{JHEP} {\bfseries 07} (2020) 073}, [\href{https://arxiv.org/abs/1912.04169}{{\ttfamily 1912.04169}}].

\bibitem{Copetti:2020dil}
C.~Copetti, A.~Grassi, Z.~Komargodski and L.~Tizzano, \emph{{Delayed deconfinement and the Hawking-Page transition}}, \href{https://doi.org/10.1007/JHEP04(2022)132}{\emph{JHEP} {\bfseries 04} (2022) 132}, [\href{https://arxiv.org/abs/2008.04950}{{\ttfamily 2008.04950}}].

\bibitem{Kim:2023sig}
S.~Kim, S.~Kundu, E.~Lee, J.~Lee, S.~Minwalla and C.~Patel, \emph{{Grey Galaxies\textquoteright{} as an endpoint of the Kerr-AdS superradiant instability}}, \href{https://doi.org/10.1007/JHEP11(2023)024}{\emph{JHEP} {\bfseries 11} (2023) 024}, [\href{https://arxiv.org/abs/2305.08922}{{\ttfamily 2305.08922}}].

\bibitem{Choi:2024xnv}
S.~Choi, D.~Jain, S.~Kim, V.~Krishna, E.~Lee, S.~Minwalla et~al., \emph{{Dual Dressed Black Holes as the end point of the Charged Superradiant instability in ${\cal N} = 4$ Yang Mills}},  \href{https://arxiv.org/abs/2409.18178}{{\ttfamily 2409.18178}}.

\bibitem{Bajaj:2024utv}
K.~Bajaj, V.~Kumar, S.~Minwall, J.~Mukherjee and A.~Rahaman, \emph{{Grey Galaxies in $AdS_5$}},  \href{https://arxiv.org/abs/2412.06904}{{\ttfamily 2412.06904}}.

\bibitem{Agarwal:2020zwm}
P.~Agarwal, S.~Choi, J.~Kim, S.~Kim and J.~Nahmgoong, \emph{{AdS black holes and finite N indices}}, \href{https://doi.org/10.1103/PhysRevD.103.126006}{\emph{Phys. Rev. D} {\bfseries 103} (2021) 126006}, [\href{https://arxiv.org/abs/2005.11240}{{\ttfamily 2005.11240}}].

\bibitem{Murthy:2020scj}
S.~Murthy, \emph{{Growth of the $\frac {1} {16}$-BPS index in 4d $N=4$ supersymmetric Yang-Mills theory}}, \href{https://doi.org/10.1103/PhysRevD.105.L021903}{\emph{Phys. Rev. D} {\bfseries 105} (2022) L021903}, [\href{https://arxiv.org/abs/2005.10843}{{\ttfamily 2005.10843}}].

\bibitem{Lezcano:2021qbj}
A.~G. Lezcano, J.~Hong, J.~T. Liu and L.~A. Pando~Zayas, \emph{{The Bethe-Ansatz approach to the $ \mathcal{N} $ = 4 superconformal index at finite rank}}, \href{https://doi.org/10.1007/JHEP06(2021)126}{\emph{JHEP} {\bfseries 06} (2021) 126}, [\href{https://arxiv.org/abs/2101.12233}{{\ttfamily 2101.12233}}].

\bibitem{Benini:2021ano}
F.~Benini and G.~Rizi, \emph{{Superconformal index of low-rank gauge theories via the Bethe Ansatz}}, \href{https://doi.org/10.1007/JHEP05(2021)061}{\emph{JHEP} {\bfseries 05} (2021) 061}, [\href{https://arxiv.org/abs/2102.03638}{{\ttfamily 2102.03638}}].

\bibitem{Cabo-Bizet:2024kfe}
A.~Cabo-Bizet and W.~Li, \emph{{Generalized Bethe expansions of superconformal indices}},  \href{https://arxiv.org/abs/2411.12018}{{\ttfamily 2411.12018}}.

\bibitem{Choi:2025lck}
S.~Choi, D.~Jain, S.~Kim, V.~Krishna, G.~Kwon, E.~Lee et~al., \emph{{Supersymmetric Grey Galaxies, Dual Dressed Black Holes and the Superconformal Index}},  \href{https://arxiv.org/abs/2501.17217}{{\ttfamily 2501.17217}}.

\bibitem{Kinney:2005ej}
J.~Kinney, J.~M. Maldacena, S.~Minwalla and S.~Raju, \emph{{An Index for 4 dimensional super conformal theories}}, \href{https://doi.org/10.1007/s00220-007-0258-7}{\emph{Commun. Math. Phys.} {\bfseries 275} (2007) 209--254}, [\href{https://arxiv.org/abs/hep-th/0510251}{{\ttfamily hep-th/0510251}}].

\bibitem{Gaiotto:2021xce}
D.~Gaiotto and J.~H. Lee, \emph{{The giant graviton expansion}}, \href{https://doi.org/10.1007/JHEP08(2024)025}{\emph{JHEP} {\bfseries 08} (2024) 025}, [\href{https://arxiv.org/abs/2109.02545}{{\ttfamily 2109.02545}}].

\bibitem{Imamura:2021ytr}
Y.~Imamura, \emph{{Finite-N superconformal index via the AdS/CFT correspondence}}, \href{https://doi.org/10.1093/ptep/ptab141}{\emph{PTEP} {\bfseries 2021} (2021) 123B05}, [\href{https://arxiv.org/abs/2108.12090}{{\ttfamily 2108.12090}}].

\bibitem{Bourdier:2015wda}
J.~Bourdier, N.~Drukker and J.~Felix, \emph{{The exact Schur index of $\mathcal{N}=4$ SYM}}, \href{https://doi.org/10.1007/JHEP11(2015)210}{\emph{JHEP} {\bfseries 11} (2015) 210}, [\href{https://arxiv.org/abs/1507.08659}{{\ttfamily 1507.08659}}].

\bibitem{Bourdier:2015sga}
J.~Bourdier, N.~Drukker and J.~Felix, \emph{{The $\mathcal{N}=2$ Schur index from free fermions}}, \href{https://doi.org/10.1007/JHEP01(2016)167}{\emph{JHEP} {\bfseries 01} (2016) 167}, [\href{https://arxiv.org/abs/1510.07041}{{\ttfamily 1510.07041}}].

\bibitem{Murthy:2022ien}
S.~Murthy, \emph{{Unitary matrix models, free fermions, and the giant graviton expansion}}, \href{https://doi.org/10.4310/PAMQ.2023.v19.n1.a12}{\emph{Pure Appl. Math. Quart.} {\bfseries 19} (2023) 299--340}, [\href{https://arxiv.org/abs/2202.06897}{{\ttfamily 2202.06897}}].

\bibitem{Liu:2022olj}
J.~T. Liu and N.~J. Rajappa, \emph{{Finite N indices and the giant graviton expansion}}, \href{https://doi.org/10.1007/JHEP04(2023)078}{\emph{JHEP} {\bfseries 04} (2023) 078}, [\href{https://arxiv.org/abs/2212.05408}{{\ttfamily 2212.05408}}].

\bibitem{Ezroura:2024wmp}
N.~Ezroura, J.~T. Liu and N.~J. Rajappa, \emph{{Analytic continuation and the giant graviton expansion}}, \href{https://doi.org/10.1007/JHEP01(2025)028}{\emph{JHEP} {\bfseries 01} (2025) 028}, [\href{https://arxiv.org/abs/2408.02759}{{\ttfamily 2408.02759}}].

\bibitem{Lee:2023iil}
J.~H. Lee, \emph{{Trace relations and open string vacua}}, \href{https://doi.org/10.1007/JHEP02(2024)224}{\emph{JHEP} {\bfseries 02} (2024) 224}, [\href{https://arxiv.org/abs/2312.00242}{{\ttfamily 2312.00242}}].

\bibitem{Eleftheriou:2023jxr}
G.~Eleftheriou, S.~Murthy and M.~Rossell\'o, \emph{{The giant graviton expansion in $AdS_5 \times S^5$}}, \href{https://doi.org/10.21468/SciPostPhys.17.4.098}{\emph{SciPost Phys.} {\bfseries 17} (2024) 098}, [\href{https://arxiv.org/abs/2312.14921}{{\ttfamily 2312.14921}}].

\bibitem{Chang:2024zqi}
C.-M. Chang and Y.-H. Lin, \emph{{Holographic covering and the fortuity of black holes}},  \href{https://arxiv.org/abs/2402.10129}{{\ttfamily 2402.10129}}.

\bibitem{Deddo:2024liu}
E.~Deddo, J.~T. Liu, L.~A. Pando~Zayas and R.~J. Saskowski, \emph{{Giant Graviton Expansion from Bubbling Geometry: Discreteness from Quantized Geometry}}, \href{https://doi.org/10.1103/PhysRevLett.132.261501}{\emph{Phys. Rev. Lett.} {\bfseries 132} (2024) 261501}, [\href{https://arxiv.org/abs/2402.19452}{{\ttfamily 2402.19452}}].

\bibitem{Gutowski:2004ez}
J.~B. Gutowski and H.~S. Reall, \emph{{Supersymmetric AdS(5) black holes}}, \href{https://doi.org/10.1088/1126-6708/2004/02/006}{\emph{JHEP} {\bfseries 02} (2004) 006}, [\href{https://arxiv.org/abs/hep-th/0401042}{{\ttfamily hep-th/0401042}}].

\bibitem{Gutowski2004}
J.~B. Gutowski and H.~S. Reall, \emph{General supersymmetric ads5 black holes}, \href{https://doi.org/10.1088/1126-6708/2004/04/048}{\emph{Journal of High Energy Physics} {\bfseries 2004} (Apr., 2004) 048–048}.

\bibitem{Kunduri2006}
H.~K. Kunduri, J.~Lucietti and H.~S. Reall, \emph{Supersymmetric multi-charge ads5 black holes}, \href{https://doi.org/10.1088/1126-6708/2006/04/036}{\emph{Journal of High Energy Physics} {\bfseries 2006} (Apr., 2006) 036–036}.

\bibitem{Chong:2005hr}
Z.~W. Chong, M.~Cvetic, H.~Lu and C.~N. Pope, \emph{{General non-extremal rotating black holes in minimal five-dimensional gauged supergravity}}, \href{https://doi.org/10.1103/PhysRevLett.95.161301}{\emph{Phys. Rev. Lett.} {\bfseries 95} (2005) 161301}, [\href{https://arxiv.org/abs/hep-th/0506029}{{\ttfamily hep-th/0506029}}].

\bibitem{Larsen:2021wnu}
F.~Larsen and S.~Lee, \emph{{Microscopic entropy of AdS$_{3}$ black holes revisited}}, \href{https://doi.org/10.1007/JHEP07(2021)038}{\emph{JHEP} {\bfseries 07} (2021) 038}, [\href{https://arxiv.org/abs/2101.08497}{{\ttfamily 2101.08497}}].

\bibitem{Larsen:2024fmp}
F.~Larsen and S.~Lee, \emph{{Supersymmetric charge constraints on AdS black holes from free fields}}, \href{https://doi.org/10.1007/JHEP09(2024)118}{\emph{JHEP} {\bfseries 09} (2024) 118}, [\href{https://arxiv.org/abs/2405.17648}{{\ttfamily 2405.17648}}].

\bibitem{Choi:2022ovw}
S.~Choi, S.~Kim, E.~Lee and J.~Lee, \emph{{From giant gravitons to black holes}}, \href{https://doi.org/10.1007/JHEP11(2023)086}{\emph{JHEP} {\bfseries 11} (2023) 086}, [\href{https://arxiv.org/abs/2207.05172}{{\ttfamily 2207.05172}}].

\bibitem{Beccaria:2023hip}
M.~Beccaria and A.~Cabo-Bizet, \emph{{Large black hole entropy from the giant brane expansion}}, \href{https://doi.org/10.1007/JHEP04(2024)146}{\emph{JHEP} {\bfseries 04} (2024) 146}, [\href{https://arxiv.org/abs/2308.05191}{{\ttfamily 2308.05191}}].

\bibitem{Choi:2019zpz}
S.~Choi, C.~Hwang and S.~Kim, \emph{{Quantum vortices, M2-branes and black holes}}, \href{https://doi.org/10.1007/JHEP09(2024)096}{\emph{JHEP} {\bfseries 09} (2024) 096}, [\href{https://arxiv.org/abs/1908.02470}{{\ttfamily 1908.02470}}].

\bibitem{Nian:2019pxj}
J.~Nian and L.~A. Pando~Zayas, \emph{{Microscopic entropy of rotating electrically charged AdS$_{4}$ black holes from field theory localization}}, \href{https://doi.org/10.1007/JHEP03(2020)081}{\emph{JHEP} {\bfseries 03} (2020) 081}, [\href{https://arxiv.org/abs/1909.07943}{{\ttfamily 1909.07943}}].

\bibitem{Amariti:2023ygn}
A.~Amariti, J.~Nian, L.~A. Pando~Zayas and A.~Segati, \emph{{Universal Cardy-Like Behavior of 3D A-Twisted Partition Functions}},  \href{https://arxiv.org/abs/2306.05462}{{\ttfamily 2306.05462}}.

\bibitem{Choi:2021lbk}
S.~Choi, S.~Jeong and S.~Kim, \emph{{The Yang-Mills duals of small AdS black holes}}, \href{https://doi.org/10.1007/JHEP07(2024)067}{\emph{JHEP} {\bfseries 07} (2024) 067}, [\href{https://arxiv.org/abs/2103.01401}{{\ttfamily 2103.01401}}].

\end{thebibliography}\endgroup

\end{document}